\documentclass[letterpaper]{article}

\usepackage{emulateapj}
\usepackage{onecolfloat}
\usepackage{apjfonts}
\usepackage{amsmath}
\usepackage{graphicx}

\newcommand{\etal}              {{et~al.\ }}

\newcommand{\aprx}	{\mbox{$\sim$}}
\newcommand{\rxte}{{\em RXTE}}
\newcommand{\xray}	{\mbox{X--ray}}
\newcommand{\xrays}	{\mbox{X--rays}}
\newcommand{\chisq}	{\mbox{$\rm\,\chi^2$}}
\newcommand{\hxt}{HEXTE}
\newcommand{\eflux}     {\mbox{$\rm\,ergs~cm^{-2}~s^{-1}$}}

\def\heaoone{{\em HEAO-1}}
\def\ginga{{\it Ginga}}

\def\degree{\hbox{$^\circ$}}
\def\cmsq{{$\rm cm^2$}}

\def\cgro{{\it CGRO}}

\newcommand{\srcnm}	{4U~0115+63}
\begin{document}

\submitted{To Appear in the Astrophysical Journal Letters}

\twocolumn[
\title{Discovery of a Third Harmonic Cyclotron Resonance Scattering Feature in the X-ray Spectrum of 4U~0115+63}
\author{W. A. Heindl, W. Coburn,  D. E. Gruber, M. R. Pelling, 
R. E. Rothschild,}
\affil{Center for Astrophysics and Space Sciences, Code 0424, University of
California, San Diego, La Jolla, CA 92093}
\author{J. Wilms, K. Pottschmidt, R. Staubert}
\affil{Institut f\"ur Astronomie und Astrophysik -- Astronomie,
University of T\"ubingen, Waldh\"auser Strasse 64, D-72076
T\"ubingen, Germany}

\authoremail{wheindl@ucsd.edu}

\begin{abstract}

We have discovered a third harmonic cyclotron resonance scattering
feature (CRSF) in observations of the recent outburst of \srcnm\ with
the \emph{Rossi X-ray Timing Explorer} (\rxte).  The spectrum in a narrow
pulse phase range shows CRSFs at $\rm 12.40^{+0.65}_{-0.35}$, $\rm
21.45^{+0.25}_{-0.38} $, and $\rm 33.56^{+0.70}_{-0.90} $\,keV.  With
centroid energy ratios to the fundamental of $\rm 1.73 \pm 0.08$ and
$\rm 2.71 \pm 0.13$, the CRSFs are not harmonically spaced.
Strong variability of the continuum and CRSFs with pulse phase
indicate a complex emission geometry near the neutron star polar cap.
In addition, one \rxte\ observation, which spanned periastron passage,
revealed a strong 2\,mHz quasi-periodic oscillation (QPO). This is
slower by two orders of magnitude than the beat-frequency QPO expected
in this system and slower by a factor of more than 5 compared with
other QPOs seen in accreting \xray\ pulsars.
\end{abstract}

\keywords{stars: individual (4U0115+63) --- stars: neutron --- \xrays:
stars}

] % twocolumn

\section{Introduction}

The transient \xray\ source \srcnm\ is an accreting \xray\ pulsar in an
eccentric 24 day orbit (\cite{bil97}) with the O9e star,
V635 Cassiopeiae (\cite{ung98}).  \xray\ outbursts have been
observed from \srcnm\ with
\emph{Uhuru} (\cite{for76}), \heaoone\ (\cite{whe79,ros79}),
\ginga\ (e.g. \cite{tam92}), and \cgro/BATSE (\cite{bil97}).

A cyclotron resonance scattering feature (CRSF) in \srcnm\ was first
noted near 20\,keV by Wheaton, \etal\ (1979) with the UCSD/MIT hard \xray\
(10-100\,keV) experiment aboard \heaoone. White, Swank \& Holt (1983) analyzed
concurrent data from the lower energy (2-50\,keV) \heaoone/A2
experiment and found an additional feature at \aprx12\,keV.  Two
outbursts of \srcnm\ were observed with \ginga, in 1990 February and
1991 April (\cite{nag91,tam92,mih95}).  The pattern of absorption
features differed dramatically in the two outbursts.  A pair of
features similar to the \heaoone\ results in the 1990 outburst gave
way to a single feature near 17\,keV in 1991 (\cite{mih98}).

We discuss here spectral and timing analyses of observations
of the 1999 March outburst (\cite{wil99,hei99}) obtained with the
\emph{Rossi X-Ray Timing Explorer} (\rxte).

\section{Observations and Analysis}

Observations were made with the Proportional
Counter Array (PCA) (\cite{jah96}) and High Energy X-ray Timing
Experiment (\hxt) (\cite{rot98}) on board \rxte.  The PCA is a set of 5
Xenon proportional counters sensitive in the
energy range 2--60\,keV with a total effective area of \aprx
7000\,$\rm cm^2$.  \hxt\ consists of two arrays of 4 NaI(Tl)/CsI(Na)
phoswich scintillation counters (15-250\,keV) totaling \aprx 1600
\cmsq. The \hxt\ alternates between source and background
fields in order to measure the background.  The PCA and \hxt\ fields
of view are collimated to the same 1\degree\ full width half
maximum (FWHM) region.

\begin{figure}
\centerline{\includegraphics[angle=0,width=3.5in]{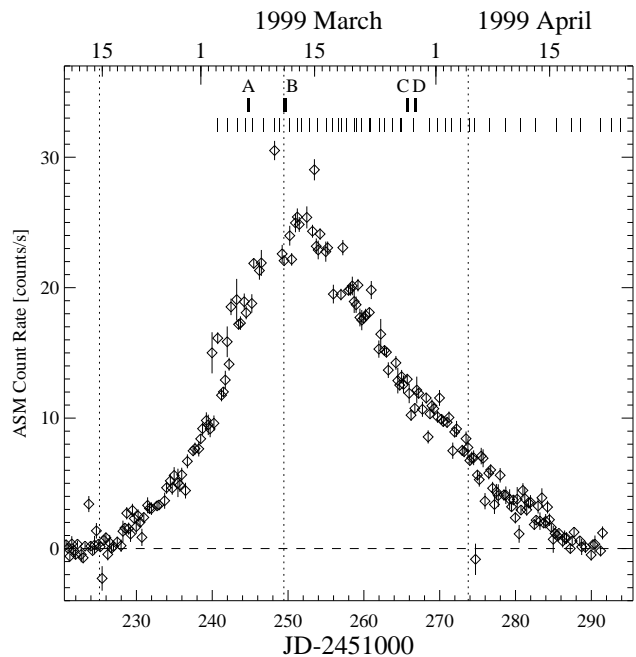}}
\figcaption{\label{f_asm} The \rxte/ASM light curve of \srcnm\ averaged in
6\,h segments.  The vertical dotted lines indicate times of periastron
passage.  The bars at the top indicate the times of the short,
public \rxte\ pointings and of the long observations (heavy bars, A--D).}
\end{figure}

Beginning on 1999 March 3, daily, short (\aprx 1\,ks) monitoring
observations were carried out.  In addition, we performed four long
pointings (duration \aprx15-35\,ks, labeled A -- D in
Fig.~\ref{f_asm}) to search for CRSFs.  
Observation B, on 1999 March 11.87-12.32, spanned
periastron passage at March 11.95 (\cite{bil97}).  Figure~\ref{f_asm}
shows the RXTE/All Sky Monitor (ASM, 1.5--12\,keV) light curve of
\srcnm\ together with the times of the pointed observations. In this
work, we concentrate on observation B.

\subsection{\label{s_anal}Spectral Analysis}

\begin{figure}
\centerline{\includegraphics[angle=0,width=4.25in]{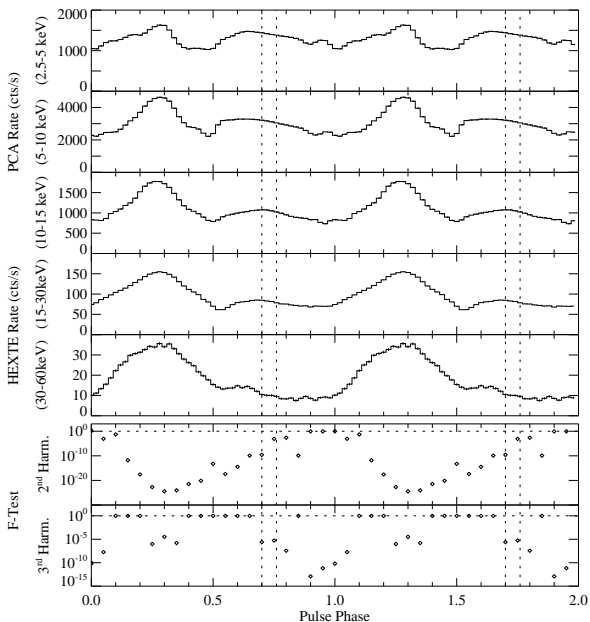}}
\figcaption{\label{f_flc} Top: The \rxte\ folded light curve of
\srcnm\ in 5 energy bands. Bottom: F-Test probability for the addition
of a second and subsequently a third CRSF to a simple continuum fit to
the HEXTE data in 20 phase bins. In both panels, points which fall on
the dashed line at $\rm 10^0$ indicate no improvement of the fit for
addition of that line.  Vertical dotted lines delineate pulse
phase 0.70--0.76 corresponding to the spectrum in Fig.~\ref{f_spec}.}
\end{figure}

The spectrum of \srcnm\ varies significantly with neutron star
rotation phase (\cite{nag91}), making fits to the average spectrum
difficult to interpret.  In order to study the evolution of the
spectrum through the pulse, we corrected photon arrival times to both
the solar system and the binary system barycenters using the ephemeris
of Bildsten, \etal\ (1997).  We then applied a Z$^2$ period search
(\cite{buc83}) to the \hxt\ data to determine the pulse period.
Figure~\ref{f_flc} shows folded light curves derived from spectra in
50 pulse phase bins for observation B, where the period was
3.614512(33)s. The folded light curve has a sharp main peak, followed
by a broader, softer second peak, similar to earlier reports
(\cite{wsh83,bil97}). In searching for CRSFs, we followed the spectral
analysis methods described in Kreykenbohm, \etal\ (1998).

Because CRSFs at \aprx12 and \aprx22\,keV are known in \srcnm\
(\cite{wsh83,nag91}), we first concentrated on the \hxt\ data where
higher harmonics might appear.  We fit a ``cut-off power law'' (a
power law times an exponential) to the \hxt\ spectra in 20 phase bins.
The reduced \chisq\ of these fits ranged from 1.3 to 12.3 (62 degrees
of freedom, ``dof''). Concentrated at the main pulse and through the
rise and peak of the second, there were significant residuals
resembling absorption features near 20--25\,keV. In the fall of the
second peak, residuals appeared between 30--40\,keV.  We then fit for
an absorption feature near 20--25\,keV, resulting in reduced \chisq s
between 0.7 and 2.0 (59 dof).  We used a simple Gaussian model for the
optical depth profile. Fig.~\ref{f_flc}, shows the result of an F-Test
for adding this line.  In the cases where no line was allowed by the
fit, the points are plotted with a value of $\rm 10^0$. Next, we
allowed a CRSF between 28--45\,keV.  This significantly improved the
fits in about half of the phase bins, including some near the main
peak where large \aprx20\,keV residuals in the initial fits masked the
presence of this line.  The corresponding F-test results are also
plotted in Fig.~\ref{f_flc}.  Although there is strong evidence for
multiple lines at other phases, the phase range 0.70--0.76 shared both
a significant \aprx20\,keV line as well as the most clearly line-like
residuals in 30--40\,keV in the no-lines fit.  We therefore chose to
concentrate on this phase in this \emph{Letter}. We plan to perform
detailed analysis of all phases and all four long (A--D) observations
in a future paper.

Next, we jointly fit the HEXTE and PCA data for phase 0.70--0.76.  To
account for uncertainties in the response matrix, 1\% systematic
errors were applied to the PCA data.  None of the continuum models
(high energy cut-off power law, Fermi-Dirac cut-off (FDCO) times a power law,
and Negative and Positive power law Exponential (NPEX); see
Kreykenbohm, \etal\ 1998) typically used for accreting pulsar spectra
provided an acceptable fit without the inclusion of absorption
features. A black body with
kT\aprx0.4\,keV and photoelectric absorption were required to describe
the data.  No Fe-K line was required.  Ultimately, it was necessary to
include CRSFs at \aprx12, \aprx21, and \aprx34\,keV in the joint
spectrum.  The fitted line parameters were insensitive to the details
of the continuum model used.  The results given here used a
Fermi-Dirac cut-off (FDCO) times a power law, given by $F(E) \propto
(1 + e^{(E - E_c)/E_f})^{-1} \times E^{-\Gamma}$.  F is the photon
flux, $\rm E_c$ the cutoff energy, $\rm E_f$ the folding energy, and
$\Gamma$ the photon index. $\rm E_c$ was fixed at zero.  An F-Test for
the significance of adding the \aprx34\,keV line to a model with only
two absorption features gave a chance probability of $\rm 10^{-17}$.

\subsection{Temporal Variability}

Along with standard Fourier techniques, we analyzed the data in the
time domain using the linear state space model (LSSM) formalism
described by K{\"o}nig \& Timmer (1997) and Pottschmidt, \etal\ (1998).  
Parameters of the LSSM are related to
dynamical timescales of the system such as oscillation periods, decay
times of damped oscillators and stochastic noise.  As shown in
Figure~\ref{f_lc}, the data are well described by a LSSM of order
2. This model, based on an auto-regressive process, is
dominated by a stochastically driven sinusoid of period 552\,s which
exponentially decays with an e-folding time of $P_{\rm fold} =
282$\,s. This short $P_{\rm fold}$ accounts for the broad QPO peak
seen in the PSD (Fig.~\ref{f_psd}).  A Kolmogoroff-Smirnoff test shows
that the difference between the data and the LSSM is purely
attributable to white noise.  The light curve is thus
consistent with a single exponentially decaying sinusoid, driven by
a white noise process.

\section{Results and Discussion}

\subsection{Spectrum and CRSFs}

\begin{figure}
\centerline{\includegraphics[angle=0,width=3.5in]{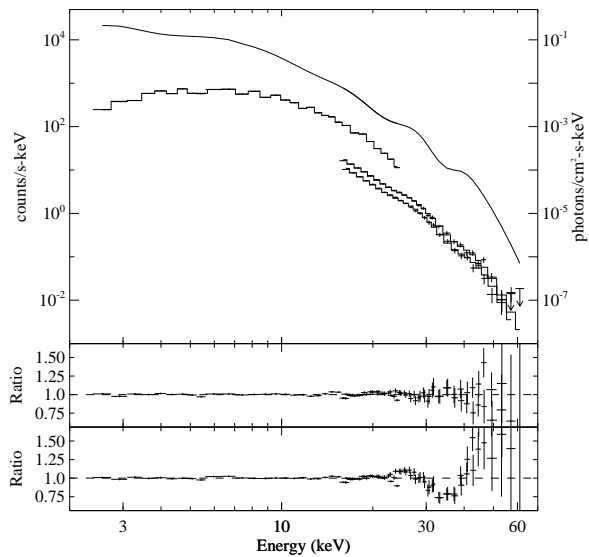}}
\figcaption{\label{f_spec} Top: The PCA and \hxt\ count spectra (crosses) for
pulse phase 0.70-0.76.  Also shown are the best fit model (histograms)
with 3 CRSFs and the inferred incident spectrum (smooth curve).
Middle: The ratio of the data to the best fit model.  Bottom:  The
ratio of the data to a model fit with only two CRSFs.  The
residuals between 30--40\,keV and the underprediction of the continuum
above 40\,keV emphasize the presence of the third line. }
\end{figure}

Fig.~\ref{f_spec} shows the best fit joint count spectrum from pulse
phase 0.70--0.76.  Also plotted is the inferred incident photon
spectrum.  Best fit parameters are given in Table~\ref{t_fit}, and the
reduced \chisq\ of the fit is 1.66 (74 dof).  This is the first time
that a fundamental and two harmonic CRSFs have been detected in a
single accreting \xray\ pulsar spectrum.  Previously, at most a
fundamental and second harmonic have been seen: 4U~1907+09
(\cite{cus98}); or suggested: Vela~X-1 (\cite{kre98}), A0535+25
(\cite{ken94}), 1E~2259+586 (\cite{iwa92}).  Those earlier
observations lacked the broad-band sensitivity of \rxte. Furthermore,
simple fits to the phase-resolved \hxt\ spectra show that the \xray\
spectrum varies rapidly with neutron star rotation.  We observe
significant variations between consecutive phase bins which cover only
2\% of the pulse phase. This suggests complex spatial variations of
conditions near the neutron star polar cap.  Since the
35\,keV CRSF is only present in about half of the pulse (predominantly
during the fall of the second, weaker pulse), and both the 22 and
35\,keV lines are only present together in 3 of 20 coarse phase bins,
averaging over large phase angles would likely have washed out the
line in the variable continuum.

The fundamental energy of 12.4\,keV implies a neutron star surface
field of $\rm 1.1\times 10^{12}(1 + z)$\,G.  Contrary to simple Landau
theory, the observed line spacing is not quite harmonic.
The ratio of the $\rm 2^{nd}$ and $\rm 3^{rd}$ harmonics to the
fundamental are $\rm 1.73 \pm 0.08$ and $\rm 2.71 \pm 0.13$
respectively. We tried fits with the $\rm
2^{nd}$ and $\rm 3^{rd}$ lines constrained to be exact
harmonics of the first, whose energy was allowed to vary. The
resulting fit was unacceptable (reduced \chisq\ of 6.7 with 76
dof). We did, however, find a reasonable fit with the 
$\rm 3^{rd}$ harmonic tied to the first but the second free to
vary. Nevertheless, an F-test comparing this fit to our best model gives a chance
improvement probability of $\rm 2 \times 10^{-4}$ for allowing the
non-integer energy ratio of the first and third lines.  
\begin{table}
\caption{\label{t_fit}  Parameters of the best fit model spectrum (see
\S \ref{s_anal}).}
\begin{minipage}{\linewidth}
\renewcommand{\thefootnote}{\thempfootnote}
\begin{tabular}{lccc} \hline \hline
 Harmonic & Energy & Width\footnote{$\sigma$ of the Gaussian optical
   depth profile} & Optical Depth \\
   & (keV)  & (keV) & \\ \hline
 1 (Fundamental) & $\rm 12.40^{+0.65}_{-0.35}$ & $\rm  3.3^{+0.1.9}_{-0.4} $ 
		& $\rm 0.72^{+0.10}_{-0.17} $\\
 2 & $\rm 21.45^{+0.25}_{-0.38} $ &$\rm 4.5^{+0.7}_{-0.9} $ 
		& $\rm 1.24^{+0.04}_{-0.06} $\\
 3 & $\rm 33.56^{+0.70}_{-0.90} $ & $\rm 3.8^{+1.5}_{-0.9}$ 
		& $\rm 1.01^{+0.13}_{-0.14} $\\ \hline
\end{tabular}
\end{minipage}
\end{table}

In addition to relativistic shifts, line energies may deviate from
harmonic for a number of reasons.  The main scattering for the
harmonics may take place in regions of different magnetic field, 
either resulting from optical depth effects in the
mound of accreting matter, or for lines primarily produced at opposite magnetic
poles.  It is interesting to note (see Fig.~\ref{f_flc}) that the
second and third harmonics are most significant in the main
and secondary pulses, respectively, possibly indicating origins at
opposite poles.

In our initial fits to the \hxt\ data alone, we observed that the
20\,keV CRSF varied in both strength and energy (by 20\%) through the
pulse phase. This was first observed with \ginga\ by Nagase, \etal\
(1991) in the 1990 Feb. outburst. The line is strongest and the energy
highest (\aprx24\,keV), on the falling edge of the main pulse, which
is similar to the behavior of the CRSFs in Cen~X-3 and 4U~1626-67
(\cite{hei99b}).  The \ginga\ data showed significant \aprx11 and
\aprx22\,keV lines at all 8 pulse phases analyzed (\cite{mih95}). The
HEXTE data find that the \aprx20\,keV line is not required just before
the rise of the main pulse (Fig.~\ref{f_flc}).  However, with the
addition of the PCA data, which constrain the continuum and
fundamental line, it is possible that this line will be required at
all phases. In any case, strong long term variability
of the lines is known (\cite{mih95}), so differences between these
results and earlier observations are not surprising.

\subsection{Temporal Variability: a 2\,mHz QPO}

\begin{figure}
\centerline{\includegraphics[angle=0,width=3.5in]{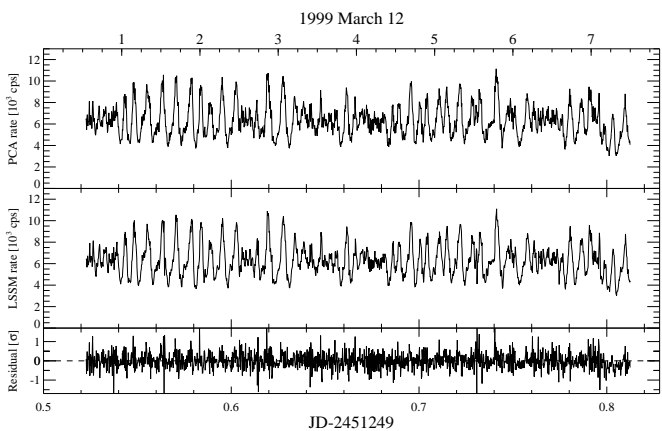}}
\figcaption{\label{f_lc} PCA (1.5-60\,keV) light curve in 16\,s bins of
observation B together with a second order linear state space model of
the variability.  The bottom panel shows the residuals of the LSSM.  }
\end{figure}

\begin{figure}[h]
\centerline{\includegraphics[angle=0,width=3.5in]{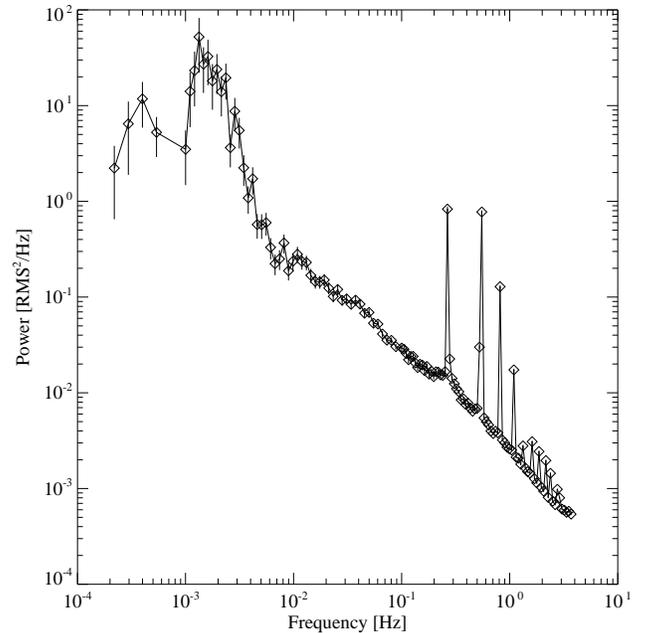}}
\figcaption{\label{f_psd}The PSD (normalized according to Miyamoto et
al. 1991) of the observation B PCA light curve. The PSD has been
rebinned such that $\rm df/f =$ 0.35, 0.1 and 0.05 in the
frequency bands $10^{-4}$--$10^{-3}$\,Hz, $10^{-3}$--$10^{-1}$\,Hz,
and $10^{-1}$--$10^{1}$\,Hz respectively. The \aprx2\,mHz QPO and
several harmonics of the 3.6\,s pulsation are clear above a 1/f
continuum. }
\end{figure}

Figures~\ref{f_lc} and~\ref{f_psd} show the PCA light curve of
observation~B and its power spectral density (PSD), respectively.
Strong variability with an \aprx500\,s period is obvious. At
frequencies above 5\,mHz, the PSD can be described by an overall power law
$\propto f^{-1}$ plus peaks at the neutron star rotational frequency
and its multiples. Some accreting pulsars have shown a QPO at the beat
frequency between the neutron star rotation and the Keplerian orbit at
the inner edge of the accretion disk (\cite{fin96}).  Using the
relations in Finger, Wilson \& Harmon (1996) with a surface B field strength
of $\rm 1.3 \times 10^{12}$\,G,
a distance of 6\,kpc (\cite{neg99}), and a total flux of 
$\rm 2.1 \times 10^{-8}$\,\eflux, the expected beat QPO
frequency is 0.8\,Hz.  Overtones of the rotational frequency confuse the
search for QPOs in this region (see Fig.~\ref{f_psd}), and
no obvious peaks apart from the pulsation were seen in the 1\,ks
short pointings or in observation B.

Below 5\,mHz, the PSD is dominated by a broad QPO feature from the
500\,s oscillation. The shape of this feature is complex and
asymmetric, it can neither be described by a Lorentzian line nor by
the superposition of two Lorentzian lines. The feature peaks at
$\sim$1.5\,mHz and has a FWHM of 1\,mHz ($Q=f_0/\Delta f \approx
1.5$). The excess power of the QPO with respect to the underlying red
noise component in the range from 1\,mHz to 4\,mHz is \aprx5\%
RMS. The \aprx500\,s period of this oscillation is longer than any
\xray\ QPO previously reported in an accreting \xray\ pulsar.  Soong
\& Swank (1989) reported a broad 0.062\,Hz QPO in \heaoone\ observations of a
flaring state in \srcnm\ that also did not fit into the beat
frequency model.

The QPO was probably present in several of the short (\aprx1\,ks)
observations as well, as there was apparent, slow variability on a
several hundred second timescale.  In observation A, the
QPO was at most present weakly, as no clear peak is evident in the
PSD. Two possible explanations for the QPO are:  1) modulation of the
accretion flow, and 2) occultation of the beam by
intervening matter in an accretion disk. It seems unlikely that the
variability is due to modulation of the accretion flow itself. The
timescales at the neutron star pole (milliseconds) and the inner edge
of the disk (seconds) are too fast.  Furthermore, the rotation period
(days) of V635Cas is too long.

We compared PCA spectra at minima of the
500\,s cycle to the spectra of the following maxima. The spectral
shape is unchanged from 2.5 -- 5\,keV, and only \aprx20\% deviations
appear at higher energies.
Because the spectrum below 5\,keV is steady through the QPO and the
flux varies by a factor of two from peak to minimum, the QPO mechanism
cannot be absorption in \emph{cold} material.  If it were,
the low energy spectrum would be highly modified by photoelectric
absorption. It is possible that Thomson scattering in ionized
matter causes the variability.  We suggest two possible mechanisms,
which both require that the accretion disk be viewed nearly edge
on. Both are consistent with a second order LSSM process.  First, an
azimuthal warp propagating around the disk could cause the ionized
disk surface to intervene in the line of sight.  In this
case, the 500\,s timescale is the time for the wave to circle the disk
(assuming a single-peaked warp).  In the second picture, the
absorption takes place in a lump in the disk which orbits at a Kepler
period of 500\,s.

\section{Summary}

We have made two discoveries in the \rxte\ observations of the
1999 March outburst of \srcnm.  The \hxt\ data have revealed for the
first time in any pulsar a third harmonic CRSF.  The line spacing
between the fundamental and second harmonic and between the second and
third harmonics are not equal, and are not multiples of the
fundamental line energy.  We have also discovered the slowest (2\,mHz)
QPO yet observed from an accreting pulsar.  It was most pronounced
during an observation spanning periastron passage of the neutron star
around its massive companion.  Based on the timescale, amplitude, and
energy spectrum of the oscillation, it is most likely due to
obscuration of the neutron star by hot accretion disk matter.  The
longest QPO previously observed had a timescale of 100\,s in SMC~X-1
(\cite{ang91}).

\acknowledgements

We thank E. Smith, J. Swank, and C. Williams-Heikkila
for rapidly scheduling the observations and supplying
the realtime data.  ASM data are provided by the \rxte/ASM teams at
MIT and at the \rxte\ SOF and GOF at NASA's GSFC.  This work was
supported by NASA grants NAS5-30720 and NAG5-7339, DFG grant Sta
173/22, and a travel grant from the DAAD.

\end{document}